\documentclass[seceq]{ptptex}

\usepackage{graphicx}
\usepackage{wrapft}

\title{ 
Gamma ray and infrared emission from the M87 jet and torus
}

\author{
Alina C. \textsc{Donea} and  Raymond J. \textsc{Protheroe}%
}

\inst{
Department of Physics and Mathematical Physics, The University of Adelaide, Adelaide, SA 5005, Australia
}

\abst{

 The existence of intrinsic obscuration of Fanaroff-Riley I
objects is a controversial topic.  M87, the nearest such object,
is puzzling in that although it has very massive central black hole
it has a relatively low luminosity, suggesting it is in a dormant
state.  Despite of its proximity to us (16 Mpc) it is not known
with certainty whether or not M87 has a dusty torus.

Infrared observations indicate that if a torus exists in M87 it
must have a rather low infrared luminosity.  Using arguments from
unification theory of active galactic nuclei, we have earlier
suggested that the inner parsec-scale region of M87 could still
harbour a small torus sufficiently cold such that its infrared
emission is dwarfed by the jet emission.  The infrared emission
from even a small cold torus could affect through photon-photon
pair production interactions the escape of 100 GeV to TeV energy
gamma rays from the central parsec of M87.  

The TeV gamma-ray flux from the inner jet of M87 has recently
been predicted in the context of the Synchrotron Proton Blazar
(SPB) model to extend up to at least 100~GeV (Protheroe, Donea,
Reimer, 2002).  Subsequently, the detection of gamma-rays above
730 GeV by the HEGRA Cherenkov telescopes has been
reported.  We discuss the interactions of
gamma-rays produced in the inner jet of M87 with the weak
infrared radiation expected from a possible dusty small-scale
torus, and show that the HEGRA detection shows that the
temperature of any torus surrounding the gamma-ray emission
region must be cooler than about 250~K.  We suggest that if no
gamma-rays are in future detected during extreme flaring activity
in M87 at other wavelength, this may be expected because of torus
heating. 

}

\begin{document}

\maketitle

\section{Introduction}

M87 is usually classified as a Fanaroff-Riley Class I (FRI) radio
galaxy having a relativistic jet \cite{Biretta1999} pointing at
$\sim 30^{\circ}$ to the line of sight\cite{Bai}.  Despite of its
proximity to us ($\sim$16 Mpc \cite{Cohen}) it is not known with
certainty whether or not M87 has a dusty torus.  However, it does
appear to have a deficiency of dust at pc scales compared to
other galaxies of its class \cite{Sparks}, and this may suggest
that the supply of dust in M87 is scarce.  Non-observations of
polarized emission from the core \cite{Zavala} suggests 
 that we see the nucleus through a
depolarizing layer which may be ionized gas filling 
 the torus.

Chandra X-ray observations of the nucleus of M87 show that the
accretion flow has low radiative efficiency \cite{DiMatteo} which
would not appear to favour a standard accretion disk
($\alpha$-disk), and therefore it is likely that the inflowing
gas settles into an advection-dominated accretion flow
(ADAF)\cite{Narayan,DiMatteo,Reynolds}.  As a consequence, the
heating of any dusty torus would be currently inefficient.
Strong mid-infrared emission was not detected from the centre of
M87\cite{Whysong, Perlman}, suggesting either that there is no
torus, or that the dust cannot settle easily into a 'standard'
well-defined large-scale torus configuration.  Corbin et
al.\cite{Corbin}, have obtained HST WFPC2 images of M87, and from
these they did not exclude the possibility of an outer radius of
a torus, if present, being smaller that 50 pc.

The existence of the intrinsic obscuration of FRIs is a
controversial topic (see Whysong \& Antonucci \cite{Whysong2003}
and references therein). Chiaberge et al.\cite{Chiaberge} suggest
that FRIs (including M87) with detected optical fluxes could lack
tori, or have geometrically very thin tori which allow the
optical emission to be seen. On the other hand, Gambill et
al. \cite{Gambill} suggested that at least some FRIs show excess
X-ray absorption, possibly a signature of a molecular torus
obscuring the core. In addition, several FRI objects show
incontestable proof of having toroidal dust structures: Centaurus
A \cite{Alexander,Whysong2003}, 3C270\cite{Barth},
3C218\cite{Sambruna,Antonucci}.
 
The presence of a broad line region in M87 is undeniable as the
Ly$\alpha$ line has observed\cite{Sankrit} a width of
3000~km~s$^{-1}$, providing direct evidence that BLR clouds are
active in M87. Either the NLR or BLR may contribute to the
observed depolarization of the radio flux \cite{Zavala}.  Donea
\& Protheroe\cite{DoneaProtheroe02b} have discussed the
coexistence of BLR and tori in active galactic nuclei (AGNs) in
the context of unification theories of AGN and their implications
for TeV emission from jets.  In this context, it is hard to
accept that M87 completely lacks a dusty component.  M87 might
still have sufficient dust around the nucleus such that its
infrared emission may affect gamma-ray escape\cite{Donea02}.  The
putative torus of M87 probably differs from a standard quasar
torus by being smaller, probably cooler, having a wide opening
angle ($\phi \ge 30^\circ$) and a large density gradient in the dust.
Such a torus would be quite different from the large tori with
external radii of $\approx 100-200$~pc assumed to populate
quasars. The limits imposed in modelling this torus come from the
fact that the infrared flux must not exceed the detected mid-IR
flux detected. In Section 3 we will discuss the geometry of the
torus in details.

Since M87 has a large central mass, and by considering aging
arguments\cite{Reynolds,Owen} it could harbour a
nucleus comprising of two or more black holes.  For a binary
black hole system a dusty torus could form with a clumpy
structure with rather diluted polar dust caps filling the space
between the jet and the equatorial belt of the torus
\cite{Zier,Biermann}.

We shall discuss the implications of the existence of such a
patchy, or clumpy, small-scale torus in M87 for observing
gamma-rays from the inner jet.  At the distance of M87 absorption
by photon-photon pair production on the intergalactic infrared
background can be neglected below 1 TeV.  We suggest that the
detection or non-detection of gamma-ray fluxes from M87 can
provide useful information about tori in FRIs.  However, if an
ADAF is present in place of a standard accretion disk, as seems
likely in M87, its radiation field may also contribute to the
optical depth for escape of gamma-rays from regions of the jet
close to the black hole.

\section{Modeling the spectral energy distribution of M87}

\begin{figure} 
\includegraphics[width=12 cm]{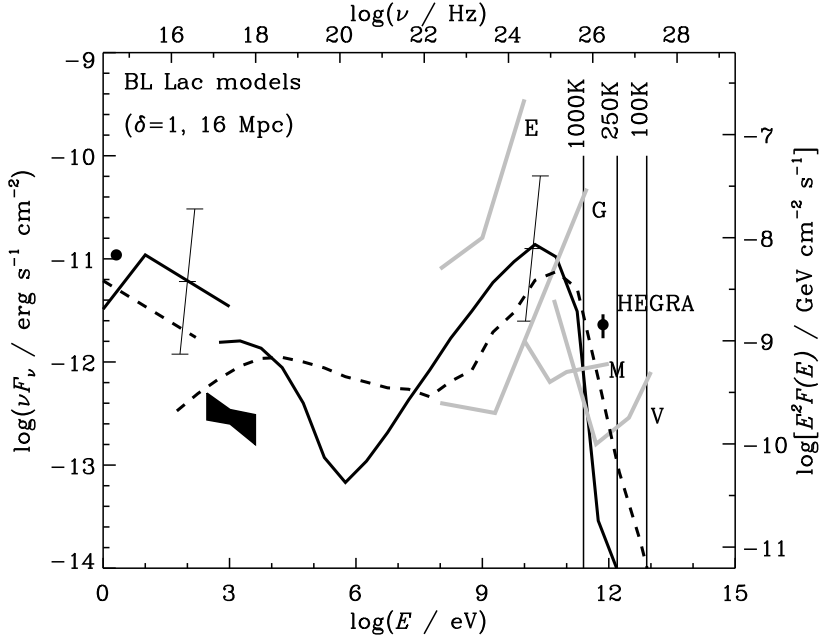} 
\caption{Predicted high energy component of the spectral energy
distribution~\protect\cite{PDR2003} (SED) of the nucleus of M87
assuming it to be a mis-aligned HBL (solid curves) or LBL (dashed
curve).  The broken power-laws on the left show the electron
synchrotron radiation which provides the target photons for
proton interactions; the curves on the right show the predicted
X-ray to gamma-ray flux.  Error bars attached to solid curves
correspond to uncertainty due to the Doppler factor of M87 which
we taken to be in the range $0.67 \le \delta \le 1.5$.  The
sensitivities of the EGRET (E), GLAST (G), MAGIC (M) and VERITAS
(V) gamma ray telescopes are indicated, as well as the HEGRA
observation~\cite{Aharonian}.  Non-simultaneous data from M87
unresolved nuclear jet emission
refs.~\protect\cite{BirettaSternHarris91,Perlman,Wilson}. Vertical
lines show the cut-off energies for photon-photon pair production
for $\gamma$-rays from the inner jet interacting with IR photons
from the assumed torus with the temperatures indicated.  }
\label{fig:1} 
\end{figure}

In this section we summarize the main results from modeling the
high energy gamma-ray emission from the inner jet of M87.  The
origin of the high-energy component of the SEDs of AGNs, starting
at X-ray or $\gamma$-ray energies and extending in some cases to
TeV-energies, is uncertain.  Using a leptonic model, Bai \&
Lee\cite{Bai} suggested that M87 could emit detectable gamma rays
with an inverse-Compton emission peak at 0.1 TeV.  In the context
of hadronic models, the spectrum of gamma-ray emission from the
nucleus of M87 has been predicted  using the SPB model by
Protheroe et al.\cite{PDR2003} by treating M87 as a mis-aligned
BL Lac object.  The SPB model \cite{MP2001a} employs hadronic
interactions in relativistic jets and the high-energy radiation
is produced through photomeson production, proton and muon
synchrotron radiation, and subsequent pair-synchrotron cascading
in the highly magnetized environment.  In the case of BL Lacs
internal photon fields (i.e. produced by synchrotron radiation
from the co-accelerated electrons at the radio to UV or X-ray
frequencies) serve as the target for pion photoproduction.
Protheroe et al.\cite{PDR2003} have shown that, M87 could be
either a high-frequency peaked (HBL) or a low-frequency peaked
(LBL) BL Lac object, and predicted the gamma-ray emission (see
Fig~\ref{fig:1}) from the nuclear jet to extend up to at least
100~GeV.  Subsequently the HEGRA Collaboration reported the
detection of gamma rays above 730 GeV with luminosity $\sim
10^{41}$~erg/s\cite{Aharonian}, consistent with that
predicted\cite{PDR2003}.  Future telescopes with higher
sensitivity (see Fig.~\ref{fig:1}) such as GLAST and MAGIC, and
possibly VERITAS, should be able to measure the spectrum in
detail.

The difference in the shapes of the high energy hump of the SED
when M87 is modelled as a mis-aligned LBL or HBL is due the
nature of the proton interactions and interactions of particles
and radiation produced as secondaries.  If M87 is modelled as a
mis-aligned LBL (dashed curve in Fig~\ref{fig:1}) pion
photoproduction losses determine the maximum proton energy, and
hence the maximum gamma-ray energy which results from the
cascades initiated by proton synchrotron radiation and pion decay
(including pion and muon synchrotron radiation).  In the case of
mis-aligned HBL, because of the lower energies of target photons,
synchrotron losses determine the maximum proton energy, and
synchrotron radiation by muons can become important (see
Protheroe et al.\cite{PDR2003} for details).

The
uncertainties in the Doppler factor of M87 (range considered
$0.66$ -- $1.6$) give rise to a much larger uncertainty in the
bolometric luminosity (indicated in the theoretical SED by the
slanted error-bars).  In both cases the predicted neutrino flux is well below
detection levels of future neutrino telescopes, however it is
possibly that the flux will increases during an extreme flare in
the LBL case.

\section{Gamma ray absorption in torus and ADAF fields}

\begin{wrapfigure}{l}{8.5cm} 
\includegraphics[width=8.5 cm,height=7 cm]{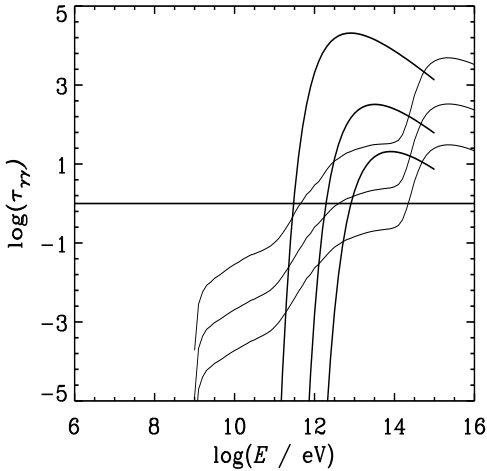} 
\caption{Optical depth for the absorption of GeV--TeV photons
travelling along the jet from distance $z$ from the black hole to
infinity: in the infrared field of a compact torus with
temperature $T=(100, 250, 1000)$~K (thick curves from bottom to
top, $z \ll 1$~pc); in the radiation field of an ADAF for
$z=(0.001, 0.01, 0.1)$~pc (thin curves from top to
bottom).  }  \label{fig:2}
\end{wrapfigure}

We calculate the photon-photon pair-production optical depth
$\tau_{\gamma\gamma}$ of GeV-TeV photons emitted by the jet of
M87 interacting with IR photons from the torus as in Donea \&
Protheroe \cite{DoneaProtheroe02b}.  In Section~1 we mentioned
that M87 could have an ADAF, and with the jet contributing little
to the heating the torus, a compact cold torus would result.  We
take the inner radius of the torus to be $R_{\rm in}\approx
1$~pc.  For an opening angle of at least $\phi \approx
30^{\circ}$, such that the inner jet could be seen from the
observer we infer a maximum height of $1.7$~pc of the inner wall
of the torus. If the torus cross section is a rectangle with the
inner edges cut way at the angle $\phi \ge 30^{\circ}$,
then the torus could be higher at radii larger than $2.5$~pc (such
that the nucleus remains unobscured). Since the $\gamma$-ray
emission is assumed to come mainly from the inner jet (distances
$z \ll 1$~pc), then it is completely imbedded in the infrared
radiation of the torus. In this case the outer radius of the
torus is irrelevant, and only the temperature of its inner wall
is important. However, if the $\gamma$-rays are produced
somewhere above the torus, then the radial extent of the dust
should be considered. We take a minimum value for the outer
radius of the torus $R_{\rm out}=2.5$~pc, and from $R_{\rm out}
\approx 2.5$~pc to 50~pc\cite{Corbin} the dust is considered to
have a very patchy configuration.  The calculation of the torus
emission follows Barvainis\cite{Barvainis} where the ultraviolet
absorption efficiency of the grains is taken to be unity. A torus
heated up to $T \sim 250$~K could account for the observed
infrared flux.

Fig~2 shows the $\gamma$-$\gamma$ opacity for $\gamma$-rays
travelling along the jet axis from $z \ll 1$~pc to infinity in
the infrared radiation of a compact torus for three temperatures
$T=(100, 250, 1000)$~K.  It can be seen that $\gamma$-rays with
energies above 10 TeV could still be absorbed even when the torus
is cold ($T=100$~K) provided the emission region lies within the
torus. 

Radiation emitted by an ADAF may also provide target photons for
photon-photon pair production.  We have taken the geometry of the
ADAF to be a disk of inner radius 6 gravitational radii and the
outer radius at 1000 gravitational radii, and the ADAF spectrum
is taken from Di Matteo et al.\cite{DiMatteo}.  The
$\gamma$-$\gamma$ opacity for $\gamma$-rays interacting with the
anisotropic ADAF radiation field is also shown in Fig~2 for three
different positions of the emitting region along the jet.  As one
can see, the TeV emission could be attenuated if the gamma-ray
emission region is sufficiently close to the black hole.

\section{Discussion}

The gamma-ray flux observed from M87 by HEGRA is for energies
above 730~GeV, the flux level is such that with the sensitivity of
the HEGRA telescopes it is impossible to determine to what
energies the spectrum continues (the predicted fluxes also not
extend TeV energies).  Nevertheless, given the HEGRA data it is
possible to say something about the nature of any torus present.
With the emission region well inside the putative torus, one
expects a cut-off in the gamma rays spectrum at an energy
determined by the torus temperature (vertical lines in Fig~1).
The fact that HEGRA has detected gamma-ray emission at 730~GeV
confirms that the torus (if it exists) should be cold, with a
temperature less than $\sim 250$~K, consistent with the
interpretation of the directly observed infrared
flux\cite{Perlman}.

M87 appears to be in a dormant state, and the situation could change
drastically during extreme flaring activity, possibly related to
spin flip during binary black hole merger\cite{Biermann}, turbulences or a sudden increase in the  accretion mass rate.
Owen et al.\cite{Owen} pointed out that M87 could display on-off
activity cycles, assuming that there could be a phenomenon which turns
off/on the central engine.  In this case the small-scale torus could
undergo periodic heating during which the torus temperature could rise
up to the sublimation temperature $\sim 1500$~K. This would present
any TeV photons produced within the {\it reactivated} torus escaping
(see the top solid curve of Fig~2 corresponding to $T=1000$~K).  Such
a reactivation of the nucleus of M87 would affect other radiation
fields making them relevant target fields for interaction of particles
and radiation within the jet.  In this case the SPB model would need
to be replaced with a different model, perhaps a proton quasar
model~\cite{PD2003} in which pion photoproduction and Bethe-Heitler
pair production dominate for energetic protons.  In this case, a
reduction in the TeV gamma ray flux would be accompanied by an
increase in the flux at other wavelength, and also to enhanced
neutrino emission, perhaps at a level observable by planned km$^3$
extensions to the AMANDA neutrino detector.  Regular monitoring of M87
in several wavelength bands would be helpful to detect the onset of
extreme flaring activity.

\section*{Acknowledgments}
This work is supported by a grant from the Australian Research Council.
A. Donea would like to acknowledge the hospitality offered by the University of Tokio in January 2003.

%


\begin{thebibliography}{99}
  
\bibitem{Biretta1999}  J.A. Biretta, in The Radio Galaxy M87,
ed. H.-J. R\" oser \& K. Meisenheimer (Springer: Berlin), (1999) 159.
\bibitem{Bai} J.M. Bai and  M.G. Lee, Astrophys. J. \textbf{549} (2001) L173

\bibitem{Cohen}J.C. Cohen et al. , \JL{Astron.\ J.\  ,119,2000,162}.

\bibitem{Sparks}W.B. Sparks, J.A. Biretta, F. Macchetto, \JL{Astrophys. J.,
473 ,1996, 254}

\bibitem{Zavala} R.T. Zavala ,G.B. Taylor, \JL{Astrophys. J.,  566,2002, 9}

\bibitem{DiMatteo} T. Di Matteo, S.W. Allen, A.C. Fabian, A.S. Wilson, Andrew J. Young, \JL{Astron.\ J.\ ,582,2003,133}

\bibitem{Narayan} R.~Narayan, I~Yi and R~Mahadevan, \JL{Nature, 375,1995,623} 

\bibitem{Reynolds} C.S.  {Reynolds} ,T. Di {Matteo}, A. C., {Fabian} , U. Hwang  and C.R. Canizares ,\JL{ MNRAS, 283, 1996,L111}

\bibitem{Whysong} D. Whysong, R. Antonucci, astro-ph/0106381

\bibitem{Perlman} E.S. Perlman, W.B. Sparks, J. Radomski, C. Packham, R.S. Fisher, R. Pina, J.A. Biretta, Astrophys. J. \textbf{561} (2001) L51


\bibitem{Corbin}M.R. Corbin , E.  O'Neil, M.  Rieke, \JL{Astron. J.,124,2002,183}

\bibitem{Whysong2003} D. Whysong, R. Antonucci, astro-ph/0207385

\bibitem{Chiaberge}M. Chiaberge,  A. Capetti,  A. Celloti,  \JL{A\&A, 349,1999, 77}.

\bibitem{Gambill} J.K. Gambill, R.M. Sambruna, G. Chartas, et al., accepted  A\&A  (2003), astro-ph/0302265

\bibitem{Alexander}D.M.~Alexander , A.~Efstathiou, J.H.~Hough, et al.,  \JL{MNRAS,310,1999,78}



\bibitem{Barth} A.J Barth, A.V. Filippenko,  Astrophys. J. \textbf{525} (1999)  673.
\bibitem{Sambruna} R.M. Sambruna, G. Chartas, M. Eracleous,  R.F. Mushotzky,\JL{Astrophys. J.,532,2000,91}


\bibitem{Antonucci}   Antonucci, R.,  Tenerife lectures, (2001) astro-ph/0103048. 




\bibitem{Sankrit}R.  Sankrit, K.R.  Sembach , \& C.R.  Canizares, \JL{Astrophys. J.,  527, 1999,733}


\bibitem{DoneaProtheroe02b} A.-C. Donea, R.J. Protheroe, \JL{ Astroph. Phys., 18,2003,377}


\bibitem{Donea02}  A.-C. Donea, R.J. Protheroe, Proceedings of the conference on Active Galactic Nuclei: from Central Engine to Host Galaxy, meeting held in Meudon, France,
    July 23-27, 2002, Eds.: S. Collin, F. Combes and I. Shlosman,astro-ph/0301433
\bibitem{Owen} F.N.  Owen, J.A. Eilek  and  N.E. Kassim \JL{Astrophys. J.,ApJ, 543,200, 611} 
\bibitem{Zier}C. Zier , P.L. Biermann ,   A\&A, 377 (2001) 23

\bibitem{Biermann} P.L.~Biermann et al., (2002), astro-ph/0211503 

\bibitem{MP2001a}  A. M\"ucke, R.J. Protheroe, \JL{Astropart. Phys,  15,2001,121}




\bibitem{Aharonian}HEGRA Collaboration: F. Aharonian, A. Akhperjanian, M. Beilicke,  submitted to A\&A (2003), astro-ph/0302155  


\bibitem{BirettaSternHarris91} J.A. Biretta, C.P. Stern, D.E. Harris, Astron. J. \textbf{101} (1991) 1632.


\bibitem{GLAST} ``Exploring nature's highest energy processes with the Gamma Ray Large Area Space Telescope'' NASA document NP-2000-9-107-GSFC, p29.,
 http://glast.gfsc.nasa.gov/resources/brochures/gsd/




\bibitem{Barvainis} V.~Barvainis, \JL{Astrophys. J., 320, 1987,537}



\bibitem{PDR2003} R.J. Protheroe,  A.-C. Donea, A. Riemer, \JL{ Astroph. Phys., to 
appear, 2003,.}

\bibitem{PD2003} R.J. Protheroe,  A.-C. Donea, ICRC, (2003), in preparation



\bibitem{Wilson} A.S. Wilson ,Y. Yang,\JL{Astrophys. J.,  568,2002,133}



\end{thebibliography}
\end{document}